\documentclass[12pt]{iopart}
\usepackage{graphicx}
\usepackage{color}
\usepackage{amssymb}

\begin{document}
\hyphenation{TiOCl TiOBr}
\title[Unoccupied electronic structure of TiOCl]{Unoccupied electronic structure of TiOCl studied by x-ray absorption near-edge spectroscopy}

\author{S~Glawion$^1$, M~W~Haverkort$^2$, G~Berner$^1$, M~Hoinkis$^1$, G~Gavrila$^3$, R~Kraus$^4$, M~Knupfer$^4$, M~Sing$^1$ and R~Claessen$^1$}
\address{$^1$ Experimentelle Physik 4, Universit\"at W\"urzburg, 97074 W\"urzburg, Germany}
\address{$^2$ Max Planck Institute for Solid State Research, 70569 Stuttgart, Germany}
\address{$^3$ BESSY, Berlin, Germany}
\address{$^4$ IFW Dresden, Postfach 270116, D-01171 Dresden, Germany}
\ead{claessen@physik.uni-wuerzburg.de}

\date{\today} 

\begin{abstract}
We study the unoccupied electronic structure of the spin-$\frac{1}{2}$ quantum magnet TiOCl by x-ray absorption near-edge spectroscopy (XANES) at the Ti $L$ and O $K$ edges. Data is acquired both in total electron and fluorescence yield mode (TEY and FY, respectively). While only the latter allows to access the unconventional low-temperature spin-Peierls (SP) phase of TiOCl the signal is found to suffer from significant self-absorption in this case. Nevertheless, we conclude from FY data that effects of the SP distortion on the electronic structure are absent within experimental accuracy. The similarity of room-temperature FY and TEY data, the latter not being obscured by self-absorption, allows us to use TEY spectra for comparison with simulations. These are performed by cluster calculations in D$_{4h}$ and D$_{2h}$ symmetries using two different codes. We extract values of the crystal-field splitting (CFS) and parametrize our results in often found notations by Slater, Racah and Butler. In all cases, good agreement with published values from other studies is found. 

(Some figures in this article are in colour only in the electronic version)
\end{abstract}

\submitto{\NJP} 
\maketitle 

\section{Introduction}

\begin{figure}[t]
  \includegraphics[width=\textwidth]{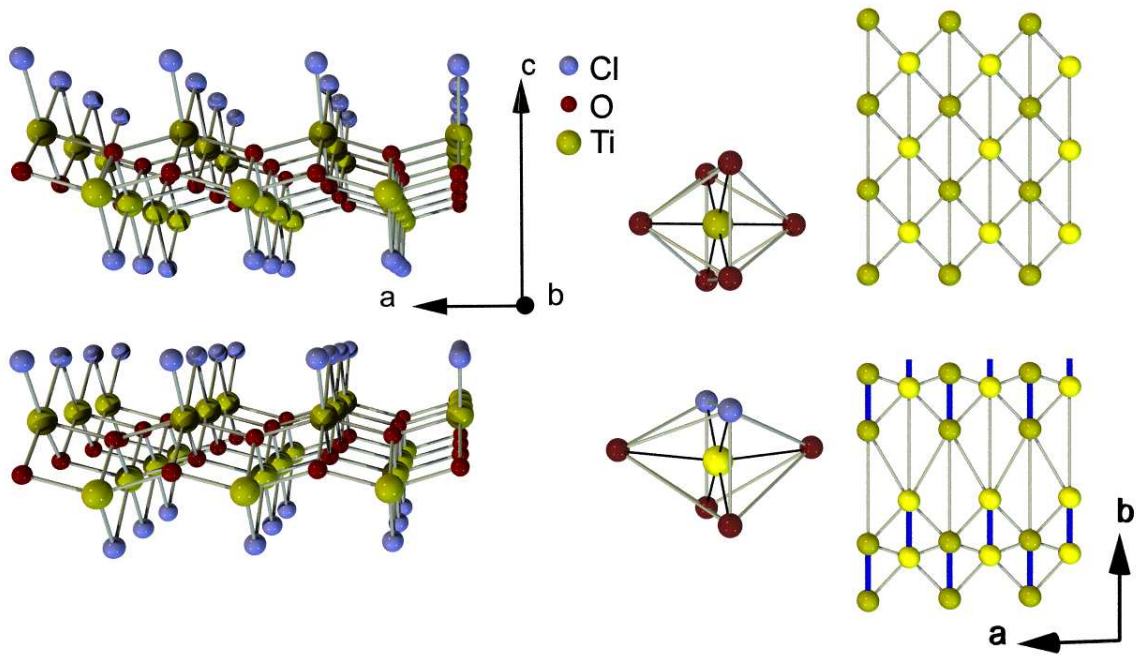}
  \caption{\label{fig:TiOCl_structure} Crystal structure of TiOCl. Left panel: Two double layers separated by van-der-Waals gaps. Center panel: Comparison of an ideal octahedral coordination (O$_h$ symmetry; top) and the actual orthorhombic coordination of Ti ions in TiOCl ($C_{2v}$ symmetry; bottom). Right panel: Projected view along the $c$ axis of the Ti sublattice. Top: Undimerized high-$T$ structure. Bottom: Dimerized spin-Peierls phase. Thick blue bonds indicate $S=0$ singlet pairs.}
\end{figure}

Low-dimensional quantum spin systems often have a rich phase diagram, especially when they involve magnetic frustration. If strong electronic correlations are also important due to partially filled $d$ or $f$ shells there will be strong interplay between spin, charge, orbital, and lattice degrees of freedom. In such systems a method like x-ray absorption spectroscopy (XAS) although it measures only the unoccupied density of states (DOS) thus gives valuable information about all degrees of freedom, not only the electronic ones. Figure~\ref{fig:TiOCl_structure} shows the crystal structure of the layered transition metal oxyhalide TiOCl. It has a spin-Peierls (SP) ground state, i.e. a dimerized phase due to dominant coupling between lattice and spin \cite{Seidel03}. This state is reached, however, in an unconventional fashion by two consecutive phase transitions \cite{Shaz05,Krimmel06,Clancy07}. That is, upon lowering the temperature a kink in the magnetic susceptibility is observed at $T_{c2}=91$\,K, and incommensurate peaks are found in x-ray diffraction (XRD). At $T_{c1}=67$\,K the susceptibility has a sharp drop and shows hysteresis, concomitant with a lock-in of the XRD peaks at commensurate $(hkl)$ positions, i.e. the dimerization of Ti ions is completed \cite{Schoenleber08}. This succession of a second-order and first-order phase transition signals the non-canonical character of the SP transition in TiOCl. While the electronic structure at elevated temperatures (room temperature and above) has been investigated intensively both by theory  \cite{Saha-Dasgupta04,Saha-Dasgupta05,Saha-Dasgupta07,Pisani07,Aichhorn09} and experiment \cite{Caimi04,Lemmens04,Lemmens05,Hoinkis05,Hoinkis07}, the low-temperature phases are not accessible by photoemission spectroscopy (PES). TiOCl is a $3d^1$ Mott insulator, and below room temperature its strongly insulating character leads to excessive charging and a loss of all spectral features in PES. Fluorescence yield (FY) x-ray absorption spectroscopy can overcome this problem since the detected photons are not susceptible to sample charging. As an analysis of FY spectra introduces further complications like e.g. self-absorption we have used total electron yield (TEY) data more extensively, as will be discussed below with appropriate caveats.

The potential of XAS at the oxygen $K$ and transition metal (TM) $L$ edges to investigate the electronic structure of strongly correlated electron systems has been shown for many materials in the past two decades \cite{Zaanen85,deGroot89,deGroot90a,Fujimori93,Okada93,Abbate93,Wende04,deGroot08}. A simple model case is when the TM ion has a $2p^6 3d^0$ valence configuration and is octahedrally coordinated by six oxygen ligands (cf. Fig.~\ref{fig:TiOCl_structure}). This gives rise to crystal-field (CF) splitting of the $3d$ orbitals into triply degenerate $t_{2g}$ and doubly degenerate $e_{g}$ levels due to the local O$_h$ symmetry. Hybridization of the TM $3d$ shell with the O $2p$ ligand orbitals allows an investigation of the unoccupied DOS of the TM ions also from the O $K$ edge spectra. An advantage of the O $K$ edge spectrum is the absence of multiplet splitting which would lead to rather complicated shapes of the absorption spectra, as seen at the TM $L$ edge. The latter, however, can be tackled by local calculations on single clusters (i.e. (TM)O$_6$) using the configuration interaction cluster model with full atomic multiplet theory. This is highly powerful when simulating $L$ edge spectra because one can include the initial and final states as well as spin-orbit coupling, multiplet splitting, and charge transfer effects on an equal footing.

The Ti ions in TiOCl lie in a strongly distorted O$_4$Cl$_2$ octahedron (cf. Fig.~\ref{fig:TiOCl_structure}), resulting in five well separated CF levels and a quenching of the orbital degree of freedom \cite{Saha-Dasgupta04,Hoinkis05,Glawion11}. The local point-group at the Ti site is $C_{2v}$, with the $C_{2}$ axis parallel to the crystal $c$ direction, pointing in between two Cl ligands. In this symmetry, the $d_{x^2-y^2}$ and the $d_{z^2}$ orbital belong to the same irreducible representation ($a_1$) and therefore can mix. It is found that the CF eigenstates in TiOCl are close to a $d_{x^2-y^2}$-like orbital pointing in the $c$ and $b$ direction for the ground state and a $d_{z^2}$-like orbital pointing in the $a$ direction for a highly excited state within the commonly used reference frame ($x=b$, $y=c$, $z=a$). Accordingly, the eigenstates can be labeled (in ascending energetic order) $a_1\, d_{b^2-c^2}=d_{x^2-y^2}$, $a_2\, d_{ab}=d_{xz}$, $b_1\, d_{ac}=d_{yz}$, $b_2\, d_{bc}=d_{xy}$, and $a_1\, d_{a^2}=d_{z^2}$, i.e. the only electron in the $3d$ shell has $d_{x^2-y^2}$ character. This provides a preferred one-dimensional hopping path along $b$ due to direct overlap of electron clouds. As the XAS final state now has two $3d$ electrons, multiplet mixing is reflected both in O $K$ edge and Ti $L$ edge spectra.

\section{Experimental and technical details}
Single crystals of TiOCl were prepared by chemical vapor transport from TiCl$_3$ and TiO$_2$ \cite{Schaefer58}. Typical sample dimensions were $1 \times 3 \times 0.2$\,mm$^3$. They were cleaved \textit{in situ} at pressures below $1\cdot 10^{-8}$\,mbar using Scotch tape to expose fresh surfaces. Experiments were performed with the MUSTANG endstation at beamline PM3 at BESSY (Berlin, Germany) with parallel recording of the FY and TEY signals. For the latter a Keithley Nanoampere meter was used to measure the drain current from the sample which was mounted in an isolated fashion. Absorption features (in particular carbon derived) of the plane grating monochromator were accounted for by dividing the TEY and FY signals by a simultaneously recorded spectrum from a gold mesh. The overall resolution amounted to 100\,meV full width at half maximum (FWHM) and the base pressure in the experimental chamber was below $5\cdot10^{-10}$\,mbar. Data was recorded in normal-incidence geometry, and the effective polarization was changed by rotating the sample stage around the beam axis. The manipulator allowed for cooling the sample down to at least 80\,K (checked by a thermocouple on the sample holder), i.e. well below the onset of the incommensurate lattice distortion at $T_{c2}=91$\,K. We compare different codes for our cluster calculations: crystal-field theory (CFT) calculations were performed using program codes originally developed by R. D. Cowan \cite{Cowan81}, P. H. Butler \cite{Butler81}, and B. T. Thole and F. M. F. de Groot \cite{deGroot90a}. As the desired symmetry is reached by successive symmetry reduction via different point groups it becomes increasingly difficult to define correct (and meaningful) CF parameters and operators for given polarizations. Thus, comparatively high tetrahedral symmetry (D$_{4h}$) with a distortion along the crystallographic $a$ axis was used to simulate XAS spectra of TiOCl. Ligand-field multiplet (LFM) calculations in D$_{2h}$ symmetry were performed with a code by M. W. Haverkort. The ligand-field theory parameters are calculated using Wannier Orbitals obtained from a density-functional theory (DFT) calculation done in the local-density approximation (LDA) with the Stuttgart LMTO code \cite{Haverkort12}.

\section{Results and discussion}
\begin{figure}[t]
  \includegraphics[width=\textwidth]{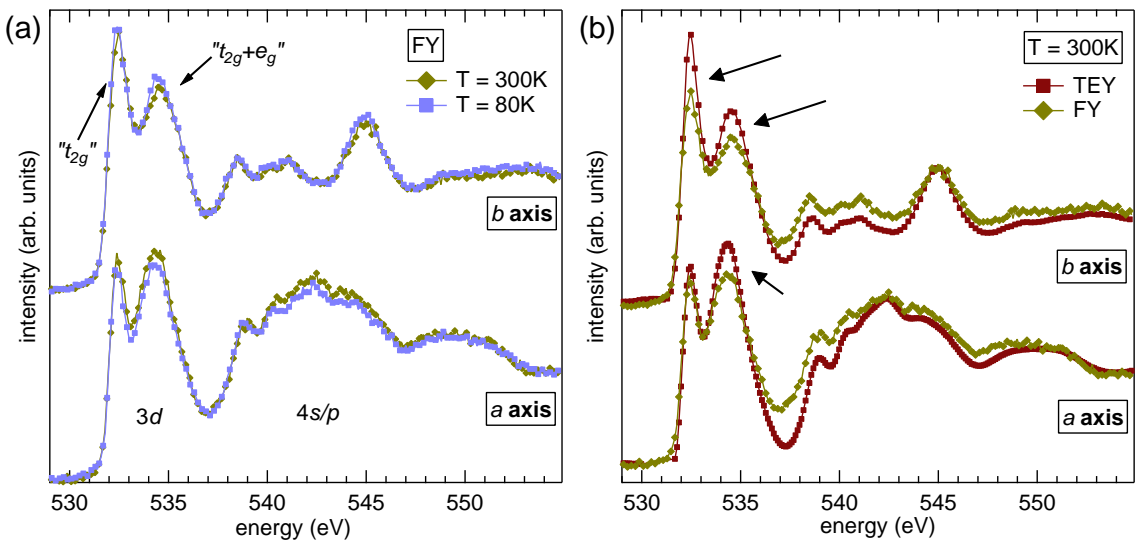}
  \caption{\label{fig:XAS_OK} O $K$ edge absorption spectra of TiOCl. Spectra with polarization along $b$ are vertically offset for clarity. (a) FY spectra at room temperature (yellow diamonds) and 80\,K (blue squares). The ranges of predominantly Ti $3d$ ($h\nu \lesssim 537$\,eV) and Ti $4s/p$ ($h\nu \gtrsim 537$\,eV) bands are indicated at the bottom. For $3d$ derived structures, the peaks of mainly $t_{2g}$ and mixed $t_{2g}$ and $e_g$ character are marked (see text for details). (b) Comparison between FY (yellow diamonds) and TEY (red squares) spectra at $T = 300$\,K. FY spectra obviously suffer from self-absorption; arrows mark prominent examples. Note that from the room-temperature TEY data along $a$, pre-edge data points between 530.0 and 531.7\,eV have been excluded as bad data points.}
\end{figure}
Figure~\ref{fig:XAS_OK} shows a complete set of O $K$ edge FY and TEY spectra taken at room temperature and at 80\,K for linear polarization both along the $a$ and $b$ axis. From the FY data one sees that no significant changes in the electronic structure take place between the high-temperature phase and the incommensurate intermediate phase ($T_{c1} < 80$\,K\,$ < T_{c2}$). Since this observation is resolution-limited, changes in the electronic structure occur obviously only on the energy scale of some 10\,meV or less. This supports the assumption that the charge degrees of freedom are barely involved in the spin-Peierls transition in this oxyhalide. From the very good correspondence between room-temperature FY and TEY spectra [cf. panel (b)] we conclude that, despite a reduced probing depth, TEY spectra reflect bulk properties of TiOCl and can be used for a detailed analysis. Thus, one circumvents possible problems from self-absorption which appear e.g. in the FY spectrum with polarization along $b$, where the intensity of the peak at lowest energy appears suppressed. The observed absorption features can roughly be assigned to O $2p$ states mixed with Ti $3d$ ($E_{ph} \approx 531-537$\,eV) or $4s/4p$ ($E_{ph} \gtrsim 537$\,eV) orbitals, respectively, and the double-peak structure in the low-energy part comes mostly from the (quasi-)\, octahedral coordination of the TM ion. In an ideal O$_h$ symmetry this peak separation would directly reflect the $t_{2g}$-$e_{g}$ CF splitting $10Dq$ \cite{deGroot90a}. Although TiOCl belongs to the point group C$_{2v}$ with considerably lower symmetry, the broadness of features as well as the illustrative value of the terms from O$_h$ symmetry justify usage of this nomenclature for qualitative arguments.

The final state of the XAS process in TiOCl involves two electrons in the $d$ shell, which basically allows for triplet and singlet configurations with both electrons in $t_{2g}$ orbitals ($t_{2g}^2$; $^3T_1$ for $S=1$; $^1T_2$, $^1E$, and $^1A_1$ for $S=0$), one in a $t_{2g}$ and one in an $e_{g}$ orbital ($t_{2g}^1 e_{g}^1$; triplet states are $^3T_{2}$ and $^3T_{1}$), as well as $t_{2g}^0 e_{g}^2$ configurations. The latter as well as $t_{2g}^1 e_{g}^1$ singlet configurations can be neglected, since they lie in energy well above all others and thus certainly do not contribute to the first two peaks. Table~\ref{tab:OKedge} gives the basic energetics of these configurations\footnote{$t_{2g}^1 e_{g}^1$ singlet configurations are included only for completeness.} in terms of $U$, $J$ and $10Dq$ \cite{Lee05}, as well as values calculated by Macovez {\it et al.} who applied an embedded cluster approach to O $K$ edge spectra of TiOBr \cite{Macovez07}. The energetic order found by these authors \cite{Macovez07} is such that the triplet configuration $^3T_1$ is at lowest energy, followed by the singlet $^1A_1$\ , and finally the $^3T_{2}$ and $^3T_{1}$ $t_{2g}^1 e_{g}^1$ triplet configurations. Due to symmetrically inequivalent O sites in the applied supercell, they could give only an energy range for the different configurations from their calculation, depending on the position of the induced core hole on a specific O site. It is well justified to assume that these results and assignments also apply to TiOCl. We thus identify the peak at $E_{ph}\approx532$\,eV as the $t_{2g}^2 e_{g}^0$ triplet configuration, while the one centered at $E_{ph}\approx534$\,eV is a superposition of $^1A_1$ singlet and $^3T_2$ triplet configurations. Calculating the eigenenergies from the relations given in the third column of table~\ref{tab:OKedge} either with $U=4.5$\,eV, $J=J_3=0.7$\,eV from GGA \cite{Saha-Dasgupta07} or $U=5.2$\,eV, $J=0.72$\,eV from best fits within our CFT calculations (last column in table~\ref{tab:OKedge}; see below for details) shows how important the attractive force between core hole and excited electrons is in x-ray absorption: in neither case can the results from the embedded cluster approach (which takes this Coulomb interaction correctly into account) be reproduced by the simple $d^2$ energy scheme. Nevertheless, it is useful in situations where no core hole is present, as is evidenced by comparison to PES studies on electron-doped TiOCl \cite{Sing11}.
\begin{table}
\caption{\label{tab:OKedge}Possible configurations, term symbols and energy eigenvalues of a $d^2$ system in O$_h$ symmetry ($t_{2g}$ threefold degenerate). The definitions of $U$ and $J$ in terms of Racah parameters can be found in table~\ref{tab:SlaterRacah}. In the second-to-last column, energy ranges (in eV) relative to the ground state ($^3 T_1$) from an embedded cluster approach to O $K$ edge spectra of TiOBr are given (taken from \cite{Macovez07}). The last column gives energies (in eV) relative to the ground state, calculated using the best-fit parameters to the Ti $L$ edge spectra from CFT calculations in D$_{4h}$ symmetry (see text for details).}
\begin{indented}
\item[]\begin{tabular}{@{}lrrrr}
\br
  configuration & term(s) &  eigenenergy$^{\rm a}$ & range in TiOBr & best-fit energy \\
  $t_{2g}^2$ triplet & $^3T_1$ &  $U - 3J_3$ & 0 & 0$^{\rm b}$\\
  $t_{2g}^2$ singlet & $^1T_2$; $^1E$ &  $U - J_3$ & -- & 1.44\\
  $t_{2g}^2$ singlet & $^1A_1$ &  $U + 2J_3$ & 0.78-1.28 & 3.40\\
\mr
  $t_{2g}^1 e_{g}^1$ triplet & $^3T_2$ &  $U - 3J_4 + 10Dq$ & 1.32-1.61 & 1.61 - 2.52\\
  $t_{2g}^1 e_{g}^1$ triplet & $^3T_1$ &  $U - 3J_1 + 10Dq$ & -- & 1.90 - 2.82 \\
  $t_{2g}^1 e_{g}^1$ singlet & $^1T_2$ &  $U - J_4 + 10Dq$ & -- & 1.65 - 3.25 \\
  $t_{2g}^1 e_{g}^1$ singlet & $^1T_1$ &  $U - J_1 + 10Dq$ & -- & 3.34 - 3.65 \\
  \br
\end{tabular}
\item[] $^{\rm a}$ Notation adopted from \cite{Lee05}.
\item[] $^{\rm b}$ The absolute value using $U$ and $J$ is 3.04\,eV.
\end{indented}
\end{table}

\begin{figure}[t]
  \centering
  \includegraphics[width=\textwidth]{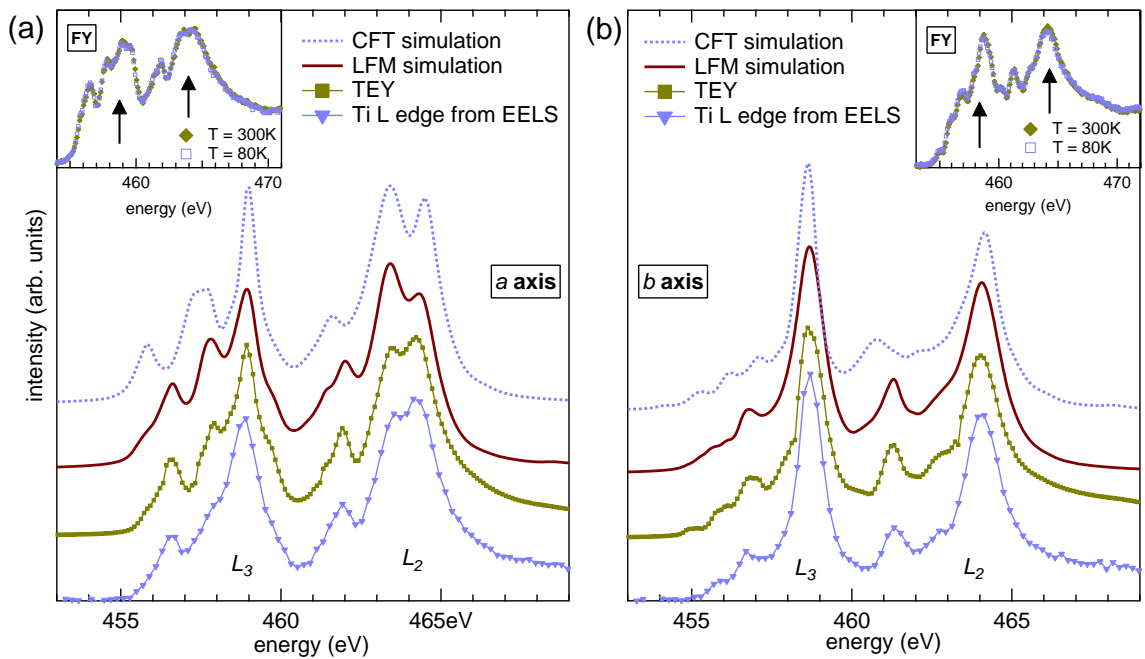}
  \caption[Ti $L$ edge absorption spectra of TiOCl]
{\label{fig:XAS_TiL} Ti $L$ edge absorption spectra of TiOCl at $T = 300$\,K with polarization (XAS) or momentum transfer (EELS) along the (a) $a$ and (b) $b$ axis. EELS data (blue triangles) provides the correct energy scale. Also shown are TEY measurements (yellow squares) and results from CFT (blue dotted curves) and LFM (red full curves) calculations in D$_{4h}$ and D$_{2h}$ symmetry, respectively (for details see text). The insets show corresponding FY spectra at room temperature (full yellow diamonds) and 80\,K (open blue squares). Self-absorption effects appear even stronger than in O $K$ edge spectra, cf. structures marked by arrows.}
\end{figure}

Upon switching the light polarization between $a$ and $b$, two major changes are observed. Firstly, the $s/p$ hybridized bands change their shape significantly. However, due to the large number of bands involved here, a simple interpretation of this behaviour is not possible, but these features serve well as a cross-check of the sample orientation. Secondly, the intensity ratio of the two $3d$ peaks is changed, making the first peak at $E_{ph}\approx532$\,eV higher than the one at $E_{ph}\approx534$\,eV for polarization along $b$. However, one cannot use simple dipole matrix arguments (i.e. selection rules) to explain this behaviour due to the mixed multiplet character of each feature. Instead, a calculation of the Ti partial DOS e.g. using density-functional theory would be required.

\begin{table}
\caption{\label{tab:XAS_Paras}Best-fit input parameters for CFT calculations in D$_{4h}$ symmetry to simulate Ti $L$ edge TEY spectra, and resulting crystal-field, Racah and Hubbard parameters. The first four rows are the input parameters to the programs, and the last four rows have been calculated from these according to known relations, cf. table~\ref{tab:SlaterRacah} \cite{Sugano70}. The values given for the Slater integrals are the atomic parameters from Hartree-Fock, which were reduced to 80\% of that value to give the best fit.$^{\rm a}$ Crystal-field splitting is given relative to the ground-state orbital. All values are given in eV.}
\begin{indented}
\item[]\begin{tabular}{@{}llll}
\br
 spin-orbit coupling & $2p$ & $3d$ & \\
 & $3.8 \pm 0.4$ & $0.0272 \pm 0.01$ & \\
\mr
 direct Slater integrals & F$^2_{pd}$ & F$^2_{dd}$ & F$^4_{pd}$\\
 & 5.581 & 10.343 & 6.499\\
\mr
 exchange Slater integrals & G$^1_{pd}$ & G$^3_{pd}$ &\\
 & 3.991 & 2.268 &\\
\mr
 crystal-field integrals$^{\rm b}$ & X$_{400}$ & X$_{420}$ & X$_{220}$ \\
 & $4.9$ & $-0.05557$ & $1.29$\\
\mr
 crystal-field parameters$^{\rm b}$ & $10Dq$ & $Dt$ & $Ds$\\
 & $1.5$ & $0.00343$ & $-0.1543$\\
\mr
 crystal-field splitting$^{\rm c}$ & $d_{xz,yz}$ & $d_{x^2-y^2}$ & $d_{z^2}$\\
 & $0.48^{\rm d}$ & $1.48$ & $2.08$\\
\mr
 Racah parameters & A & B & C\\
 & 3.559 & 0.101 & 0.413\\
\mr
 Hubbard parameters & $U$ & $J (=J_3)$ & \\
 & 5.6 & 0.74 &\\
\br
\end{tabular}
\item[] $^{\rm a}$ It is well-known that this reduction mimics the Slater integrals in actual solids \cite{deGroot08}.
\item[] $^{\rm b}$ Apply only to CFT calculations.
\item[] $^{\rm c}$ These resemble the values observed in resonant inelastic x-ray scattering (RIXS) \cite{Glawion11}. The latter were used as initial guesses for the crystal-field parameters in CFT calculations. In LFM calculations, they were fixed input instead of fit parameters.
\item[] $^{\rm d}$ This is the weighted average of the $d_{xz}$ and $d_{yz}$ CF splitting seen in RIXS \cite{Glawion11}.
\end{indented}
\end{table}

Figure~\ref{fig:XAS_TiL} presents measurements of the Ti $L$ edge for polarization along $a$ and $b$ from different methods. The EELS data was recorded with a dedicated spectrometer \cite{Fink89} using a primary beam energy of 172\,keV and an energy resolution of $\Delta E \sim 270$\,meV at $T=300$\,K. The sample preparation for these measurements and further experimental details are described in \cite{Kraus10}. The EELS energy scale was used to calibrate the photon energy scale for the XAS data. The insets show FY spectra both at $T=300$\,K and $T=80$\,K, which appear to suffer even more from self-absorption than the O $K$ edge spectra in figure~\ref{fig:XAS_OK}. No differences can be observed between the high-$T$ and low-$T$ phases, similar to the O $K$ edge. As discussed before, we used TEY data to compare with our simulations, as they show finer structures and need not be corrected for self-absorption. Results from calculations within CFT in D$_{4h}$ symmetry and from LFM theory in D$_{2h}$ symmetry are also shown in the figure. All theoretical spectra have been broadened by a Gaussian with FWHM of 0.1\,eV to account for the experimental resolution, and Lorentzians with FWHM of 0.4\,eV and 0.2\,eV for the $L_2$ and $L_3$ edges, respectively, to account for lifetime broadening. Different values for the two edges are justified by the fact that there is an additional Auger decay channel available at the $L_2$ edge, leading to enhanced broadening \cite{deGroot90a}. The best-fit parameters for the CFT calculations are given in table~\ref{tab:XAS_Paras}. Note that $d_{xz}$ and $d_{yz}$ are degenerate in D$_{4h}$ symmetry, and the value of the CF splitting given in the table thus can be understood as their weighted average.

Comparing first the CFT calculations (blue dotted curves) with the experimental spectra one realizes that although the basic shape in terms of number of peaks, relative intensity and polarization dependence can be reproduced, they appear stretched energetically compared to experiment. Such stretching can be straightforwardly attributed to the omission of charge-transfer effects (i.e. hybridization with the ligands) when one looks at the spectra calculated within LFM (dotted curve). It is well known that these effects lead to a contraction of multiplet structures if the charge-transfer energy $\Delta$ ($=5.0$\,eV for the shown spectra) is positive, and produce additional small satellite structures \cite{Okada93}. Also, the degeneracy between $d_{xz}$ and $d_{yz}$ is lifted in these calculations due to the lower D$_{2h}$ symmetry.\footnote{In D$_{2h}$ one can define two additional crystal-field parameters \cite{Haverkort05}. While $Du$ directly describes the splitting between $d_{xz}$ and $d_{yz}$, the parameter $Dv$ gives a mixing (and thus additional splitting) of $d_{z^2}$ and $d_{xy}$ orbitals.} Nevertheless, transmuting the fit parameters from the different methods into the on-site Coulomb repulsion $U$ and exchange interaction $J$ ($= J_3$ for the cases relevant to TiOCl \cite{Lee05}) one finds good agreement with typical values needed for reasonable results from density functional calculations \cite{Saha-Dasgupta07}.

In order to arrive at the LFM spectra, a tight-binding representation of the band structure on a basis of Ti-$d$, O-$p$ and Cl-$p$ Wannier orbitals was calculated using the $N$MTO approximation \cite{Anderson00}. Starting from these Wannier orbitals we defined local correlated Ti orbitals and uncorrelated ligand orbitals. The ligand orbitals were obtained as a Ti-centered linear combination of the Wannier orbitals centered at the O and Cl site. More details on this procedure can be found elsewhere \cite{Haverkort12}. For systems with a local cubic point group there is only a single ligand shell. In our case, the hybridization of the first ligand shell with the second turned out to be non-negligible and is included. For each $d$ orbital there are two ligand orbitals, labeled $L^{(1)}$ and $L^{(2)}$. The $d$ shell has interaction with the $L^{(1)}$ shell, which in turn interacts with the $L^{(2)}$ shell. The non-spherical part of the interaction found in this way is given by the following irreducible-representation-dependent one-particle Hamiltonian:
\begin{equation}
H_{a_2}=\left(
\begin{array}{c|r@{.}l r@{.}l r@{.}l}
&\multicolumn{2}{c}{d_{xz}}&\multicolumn{2}{c}{L^{(1)}_{xz}}&\multicolumn{2}{c}{L^{(2)}_{xz}}\\
\hline
d_{xz}      & -0&461 &  1&862 &\multicolumn{2}{c}{}\\
L^{(1)}_{xz}&  1&862 & -0&808 &  1&054 \\
L^{(2)}_{xz}&\multicolumn{2}{c}{}&  1&054 &  1&592
\end{array}
\right)
\end{equation}
\begin{equation}
H_{b_1}=\left(
\begin{array}{c|r@{.}l r@{.}l r@{.}l}
&\multicolumn{2}{c}{d_{yz}}&\multicolumn{2}{c}{L^{(1)}_{yz}}&\multicolumn{2}{c}{L^{(2)}_{yz}}\\
\hline
d_{yz}      &  0&013 &  2&060 &\multicolumn{2}{c}{}\\
L^{(1)}_{yz}&  2&060 & -0&399 &  1&084 \\
L^{(2)}_{yz}&\multicolumn{2}{c}{}&  1&084 &  0&929
\end{array}
\right)
\end{equation}
\begin{equation}
H_{b_2}=\left(
\begin{array}{c|r@{.}l r@{.}l r@{.}l}
&\multicolumn{2}{c}{d_{xy}}&\multicolumn{2}{c}{L^{(1)}_{xy}}&\multicolumn{2}{c}{L^{(2)}_{xy}}\\
\hline
d_{xy}      &  0&325 &  2&529 &\multicolumn{2}{c}{}\\
L^{(1)}_{xy}&  2&529 &  1&075 &  1&509 \\
L^{(2)}_{xy}&\multicolumn{2}{c}{}&  1&509 &  0&562
\end{array}
\right)
\end{equation}
\begin{equation}
H_{a_1}=\left(
\begin{array}{c|r@{.}l r@{.}l r@{.}l r@{.}l r@{.}l r@{.}l}
&\multicolumn{2}{c}{d_{x^2-y^2}}&\multicolumn{2}{c}{d_{z^2}}&\multicolumn{2}{c}{L^{(1)}_{x^2-y^2}}&\multicolumn{2}{c}{L^{(1)}_{z^2}}&\multicolumn{2}{c}{L^{(2)}_{x^2-y^2}}&\multicolumn{2}{c}{L^{(2)}_{z^2}}\\
\hline
d_{x^2-y^2}      & -0&564 & -0&081 &  1&528 & -0&123 &\multicolumn{2}{c}{}&\multicolumn{2}{c}{}\\
d_{z^2}          & -0&081 &  0&713 & -0&123 &  3&273 &\multicolumn{2}{c}{}&\multicolumn{2}{c}{}\\
L^{(1)}_{x^2-y^2}&  1&528 & -0&123 & -0&062 & -0&103 &  1&083 &  0&110 \\
L^{(1)}_{z^2}    & -0&123 &  3&273 & -0&103 &  0&194 &  0&110 &  0&910 \\
L^{(2)}_{x^2-y^2}&\multicolumn{2}{c}{}&\multicolumn{2}{c}{}&  1&083 &  0&110 &  0&930 & -0&086 \\
L^{(2)}_{z^2}    &\multicolumn{2}{c}{}&\multicolumn{2}{c}{}&  0&110 &  0&910 & -0&086 &  0&908
\end{array}
\right)
\end{equation}

The LMTO potential used to obtain the downfolded Wannier orbitals and tight-binding representation of the band structure was calculated using an atomic sphere approximation (ASA). Although reasonably accurate, the non-spherical part of the Madelung potential is missing within this potential. We included this as a set of ad-hoc parameters, fitted to the $d-d$ excitation eigenenergies as obtained from the resonant inelastic x-ray spectra of TiOCl \cite{Glawion11}. The resulting additional potentials are:
\begin{equation}
\begin{array}{ccccc}
d_{x^2-y^2}&d_{xz}&d_{yz}&d_{xy}&d_{z^2}\\
-0.036&-0.156& 0.144&-0.056& 0.104
\end{array}
\end{equation}
Within the LFM calculations the center of gravity of the $d^1$ configuration is taken to be at the energy 0. The $d^2 \underline{L}$ configuration has its average energy at $\Delta=5.0$\,eV, whereby $\underline{L}$ can either be a hole in the first or the second ligand shell. The $d^3 \underline{L}^2$ configuration has its average energy at $2 \Delta + U_{dd}$ with $U_{dd}=4.0$\,eV, and so forth.

\section{Conclusions}
We study the unoccupied electronic structure of the frustrated quantum magnet TiOCl using Ti $L$ and O $K$ edge x-ray absorption near-edge spectroscopy. From total electron yield and fluorescence yield data we conclude that the low-temperature spin-Peierls distortion has no influence on the electronic structure within experimental accuracy. Comparing our data with cluster calculations in D$_{4h}$ and D$_{2h}$ symmetries using different codes we can extract values for the crystal-field splitting which are in line with other work, and parametrize our results in widely found notations by Slater, Racah and Butler. Values of the extracted electronic parameters like $U$ and $J$ are in very good agreement with those from advanced methods like e.g. density-functional theory.

\ack 
We thank M Sperling and R Mitdank from the MUSTANG staff at BESSY for technical support during beamtime. Useful discussions with E Goering and F M F de Groot are gratefully acknowledged. This work was supported by the German Science Foundation (DFG) through grants No. CL-124/6-1 and CL-124/6-2.

\appendix
\section*{Appendix}
\setcounter{section}{1}

\begin{table}
\caption{\label{tab:SlaterRacah}Cross-relationships between normalized and unnormalized Slater-Condon parameters (SCP), Racah parameters \cite{Racah42}, and parametrization of direct and exchange Coulomb integrals in terms of these parameters (notation $J_i$ according to \cite{Lee05}).}
\begin{indented}
\item[]\begin{tabular}{@{}llll}
\br
  & SCP & normalized SCP & Racah\\
  &&&\\
  & $F^0$ & $F_0=F^0$ & $A=F^0-\frac{49}{441}F^4$\\
  SCP & $F^2$ & $F_2=\frac{1}{49}F^2$ & $B=\frac{1}{49}F^2-\frac{5}{441}F^4$\\
  & $F^4$ & $F_4=\frac{1}{441}F^4$ & $C=\frac{35}{441}F^4$\\
  &&&\\
\mr
  & $F^0=F_0$ & $F_0$ & $A = F_0 - 49F_4$\\
  normalized SCP & $F^2=49F_2$ & $F_2$ & $B=F_2-5F_4$\\
  & $F^4=441F_4$ & $F_4$ & $C=35F_4$\\
  &&&\\
\mr
  & $F^0=A-\frac{7}{5}C $ & $F_0=A-\frac{7}{5}C$ & $A$\\
  Racah  & $F^2=49B-7C$ & $F_2=B-\frac{1}{7}C$ & $B$\\
  & $F^4=\frac{441}{35}C$ & $F_4=\frac{1}{35}C$ & $C$\\
  &&&\\
\mr
  $U$ & $F^0+\frac{4}{49}F^2+\frac{36}{441}F^4$ & $F_0+4F_2+36F_4$ & $A+4B+3C$\\
\mr
  $U' = U - 2 J_H$ & $F^0-\frac{2}{49}F^2-\frac{4}{441}F^4$& $F_0-2F_2-4F_4$ & $A-2B+C$\\
\mr
  $J_1$ & $\frac{35}{441}F^4$ & $35F_4$ & $C$\\
\mr
  $J_3 (\approx J_H)$ & $\frac{3}{49}F^2+\frac{20}{441}F^4$ & $3F_2+20F_4$ & $3B+C$\\
\mr
  $J_4$ & $\frac{4}{49}F^2+\frac{15}{441}F^4$& $4F_2+15F_4$ & $4B+C$\\
\br
  \end{tabular}
\end{indented}
\end{table}

\begin{table}
\caption{\label{tab:CrystalFieldParas}Parametrization of the crystal-field levels in D$_{4h}$ using the crystal-field parameters $Dq$, $Ds$ and $Dt$, and those defined by Butler \cite{Butler81}.}
\begin{indented}
\item[]\begin{tabular}{@{}llll}
\br
 orbital & symmetry & energy in $D$ terms & energy in $X$ terms (Butler)\\
\mr
 $d_{x^2-y^2}$ & $b_{1g}$ & $6Dq+2Ds-Dt$ & $\frac{1}{\sqrt{30}}X_{400}-\frac{1}{\sqrt{42}}X_{420}-\frac{2}{\sqrt{70}}X_{220}$\\
\mr
 $d_{z^2}$ & $a_{1g}$ & $6Dq-2Ds-6Dt$ & $\frac{1}{\sqrt{30}}X_{400}+\frac{1}{\sqrt{42}}X_{420}+\frac{2}{\sqrt{70}}X_{220}$\\
\mr
 $d_{xy}$ & $b_{2g}$ & $-4Dq+2Ds-Dt$ & $-\frac{2}{3\sqrt{30}}X_{400}+\frac{4}{3\sqrt{42}}X_{420}-\frac{2}{\sqrt{70}}X_{220}$\\
\mr
 $d_{xz},d_{yz}$ & $e_{g}$ & $-4Dq-Ds+4Dt$ & $-\frac{2}{3\sqrt{30}}X_{400}-\frac{2}{3\sqrt{42}}X_{420}+\frac{1}{\sqrt{70}}X_{220}$\\
\br
  \end{tabular}
  \end{indented}
\end{table}

\section*{References}
\bibliographystyle{unsrt}


\end{document}